\documentclass[a4paper,twocolumn,floatfix]{revtex4}
\usepackage{graphicx}

\begin{document}

\title{Evolution of the conductivity of potassium-doped pentacene films}

\author{M. F. Craciun}
\author{S. Rogge}
\author{A. F. Morpurgo}

\affiliation{Kavli Institute of Nanoscience, Delft University of
Technology, Lorentzweg~1, 2628\,CJ Delft, The Netherlands}

\begin{abstract}
We have investigated the evolution of the temperature-dependent
conductivity of electron-doped pentacene (PEN) thin films, in
which the electron density is controlled by means of potassium (K)
intercalation. We find that the conductivity is high and exhibit
metallic temperature dependence in a broad range of concentrations
up to approximately 1 K/PEN. At this value, charge transfer from
potassium to PEN saturates, leaving the lowest unoccupied
molecular orbital at half filling. This causes a sharp drop of the
conductivity, concomitantly with a re-entrance into an insulating
state. Our observations are consistent with the occurrence of a
Mott-Hubbard insulating state driven by strong electron-electron
interaction in agreement with theoretical predictions.

\end{abstract}

\maketitle

Pentacene (PEN) is a conjugated molecule very well-known in the
field of plastic electronics for its use in high-mobility organic
thin-film \cite{FET}, whose electronic properties are being
investigated in great depth. One aspect that has attracted
particular attention is the possibility to control the electrical
conductivity by means of chemical doping. Doping is achieved
through the inclusion of iodine \cite{Minakata1,Minakata2,Iwasa}
and alkali atoms
\cite{Minakata3,Matsuo1,Matsuo2,electrochemistry}, that
intercalate in between the planes of the molecular films -as shown
by many different structural studies
\cite{Minakata1,Minakata2,Iwasa,Minakata3,Matsuo1,Matsuo2,electrochemistry,Matsuo3,bandstructure}-
and donate either holes (iodine) or electrons (alkali) to the
molecules. For iodine doping \cite{Minakata1,Minakata2,Iwasa}
earlier, and very recently also for rubidium doping
\cite{Matsuo3}, it has been found that the conductivity of
pentacene films can become large ($\approx$ 100 S$\cdot$cm$^{-1}$)
and exhibit a metallic temperature dependence. However, very
little is currently known about the doping dependence of the
conductivity and the microscopic nature of electrical conduction
in doped pentacene.

Here, we address these issues by investigating the evolution of
the temperature-dependent conductivity of potassium-intercalated
PEN thin films, with increasing doping concentration. We find that
the conductivity is high and exhibit metallic temperature
dependence for potassium concentrations between approximately 0.1
K/PEN and 1 K/PEN. When the amount of potassium is increased
beyond 1 K/PEN, a sharp drop of the conductivity is observed,
concomitantly with a re-entrance into an insulating state. The
analysis of the experimental data show that at high doping the
conductivity cannot be described in terms of independent electrons
filling the molecular band originating from the lowest unoccupied
molecular orbital (LUMO). Rather, our observations are consistent
with the occurrence of a Mott-Hubbard insulating state driven by
strong electron-electron interaction, in agreement with recent
theoretical predictions \cite{electronic-corelations}.\\

\begin{figure}[h]
  \centering
  \includegraphics[width=1\columnwidth]{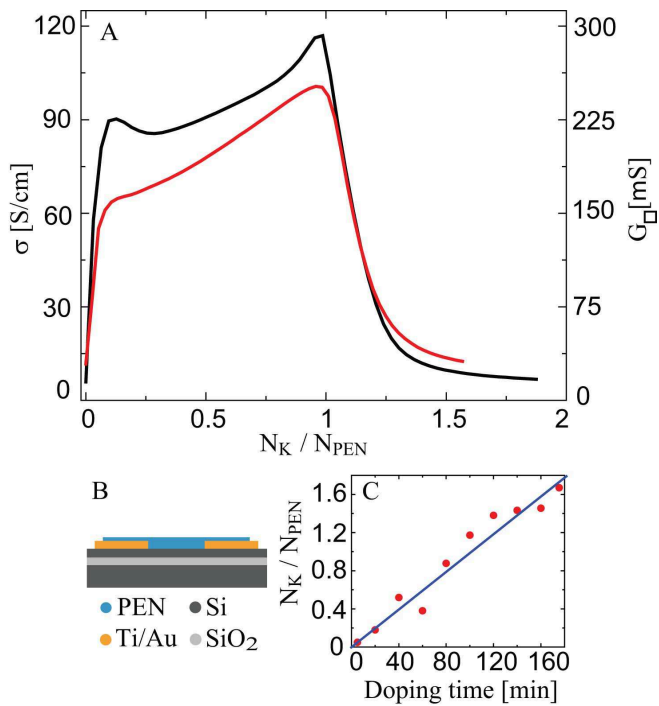}
\caption{(A) Conductivity measured at room temperature as a
function of potassium concentration for two PEN films. (B)
Schematic layout (side view) of a pentacene film on a SOI
substrate with prefabricated electrodes (not to scale). (C) Doping
calibration performed by means of RBS.}
  \label{fig:fig1}
\end{figure}

Our experiments are carried out on pentacene thin films ($\sim$
25nm thick) thermally evaporated from a Knudsen cell on an
H-terminated Si surface. The deposition of PEN films, their
doping, and the temperature-dependent transport measurements are
carried out \textit{in-situ} in a single ultra-high (UHV) vacuum
system with a base pressure $< 5 \times 10^{-11}$\,mbar. The
substrate was kept at room temperature during the evaporation of
pentacene. The film thickness was determined ex-situ using an
atomic force microscope (AFM). Prior to the film deposition,
as-purchased molecules were purified by means of physical vapor
deposition in a temperature gradient in the presence of a stream
of argon as described in ref. \cite{PENpurification}. The
substrates were prepared from a high resistive
silicon-on-insulator (SOI) wafer, consisting of a top 2$\mu$m
silicon layer electrically insulated from the Si substrate by
1$\mu$m-thick SiO$_{2}$ buried layer (see Fig.\ref{fig:fig1}B).
The electrodes (Ti 10nm thick / Au 50nm thick) were deposited on
SOI before loading the substrates in the UHV system. The
H-terminated Si surface was obtained by etching the SOI surface in
HF solution and rinsing in de-ionized water. The use of SOI
enables the growth of highly ordered films, while minimizing the
parallel conduction through the substrate. The high quality of
these films is essential, as the worst structural quality of films
deposited on insulating surfaces (e.g., glass) results in
considerably poorer electrical properties. Doping is achieved by
exposing the films to a constant flux of K atoms generated by a
current-heated getter source. The potassium concentration is
determined by means of an elemental analysis performed on
pentacene films doped at several doping levels, using ex-situ
Rutherford backscattering technique (RBS). As shown in
Fig.\ref{fig:fig1}C, the ratio of K atoms to PEN molecules is
increasing linearly with increasing the doping time, as expected.

The doping dependence of the conductivity (doping curve) of PEN
films is shown in Fig.\ref{fig:fig1}A. Despite small differences
in the doping curves of different samples (illustrated by the two
curves in the figure), the overall trend is very similar for all
the (more than 40) films measured. The conductivity rapidly
increases with doping up to a value of $\sigma \sim 100 S/cm$, it
remains high in a broad range of concentrations, and it eventually
decreases to the level observed for the pristine material
(over-doped state) for potassium concentrations higher than 1
K/PEN. We note that the observed suppression of the conductivity
at high doping exclude the possibility that the observed
conduction is due to experimental artifacts, such as, for
instance, the formation of a potassium layer on top of the
pentacene film.\\

\begin{figure}[ht]
  \centering
  \includegraphics[width=0.9\columnwidth]{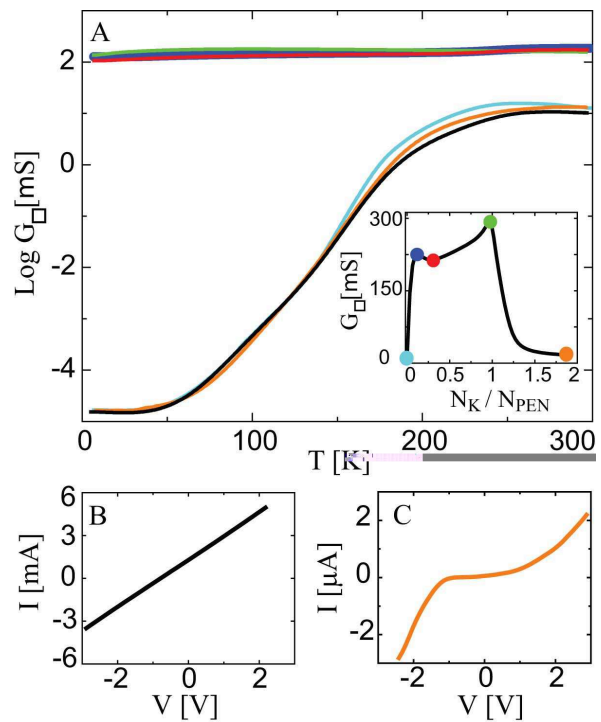}
\caption{(A) Temperature dependence of the square conductance of
pentacene films measured for the undoped, the highly conductive
and the over-doped states. The colored dots in the inset of (A)
indicate the doping level at which the temperature dependence
curves of the corresponding colors have been measured. The
temperature dependence of the SOI substrate (in black) is also
shown. Low temperature (5K) measurements of the I-V
characteristics of films in the highly conducting and the
over-doped states are shown in (B) and (C), respectively. }
  \label{fig:Tdep}
\end{figure}

To investigate the nature of electrical conduction, we have
measured the temperature dependence of the conductivity at
different doping levels (Fig.\ref{fig:Tdep}). For undoped
materials, the conductivity of the PEN film is very low, the
measurements are dominated by transport through the 2 $\mu$m thick
Si layer of the SOI substrate. The observed temperature dependence
is insulating, as expected. When the films are doped into the
highly conductive state (i.e. for K/PEN in between 0.1 and 1) the
conductance of the doped films remains high when lowering the
temperature, down to the lowest temperature reached in the
experiments ($\sim$ 5 K). This demonstrates the occurrence of
metallic conduction. In the over-doped state (for $K/PEN >1$) the
conductance is again dominated by the substrate and decreases very
rapidly with decreasing temperature, i.e. over-doped films are
insulating. To further verify the metallic nature of the different
doping states, we have measured the low-temperature (5K) I-V
characteristics at different doping levels. At potassium
concentrations between 0.1 and 1 K/PEN, the films exhibit linear
I-V characteristics (Fig.\ref{fig:Tdep}B), as expected for a
metal, whereas over-doped films show a strong non-linearity and
virtually no current when the applied bias is less than 1 V
(Fig.\ref{fig:Tdep}C), signaling a non conducting state.\\

\begin{figure}[h]
  \centering
  \includegraphics[width=1\columnwidth]{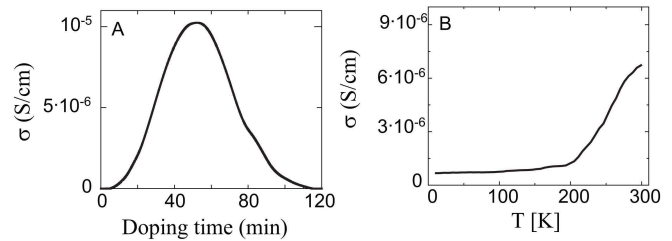}
\caption{(A) Conductivity measured at room temperature as a
function of doping time for films of pentacene deposited on glass.
(B) Temperature dependence of the conductivity for pentacene films
grown on glass measured into the optimally doped state.}
  \label{fig:glass}
\end{figure}

Earlier investigations \cite{Minakata3,Matsuo1,Matsuo2} had failed
to find a metallic behavior in alkali-doped pentacene films. The
difference between our and early results originates from the high
quality of the films that can be achieved on a hydrogen-terminated
Si surface. To clarify this important technical point,
Fig.\ref{fig:glass}A illustrates the doping dependence of films
deposited onto glass substrates.\\

The maximum conductivity (optimally doped state) is several orders
of magnitude lower than the conductivity measured for films
deposited on a Si surface. In addition, the conductivity of
potassium-doped films on glass substrates was always observed to
decrease rapidly with lowering temperature, i.e. the films are
insulating (see Fig.\ref{fig:glass}B). Both the magnitude and the
temperature dependence of the conductivity that we measured on
glass substrates compares well to results reported in earlier
work.\\

\begin{figure}[ht]
  \centering
  \includegraphics[width=1\columnwidth]{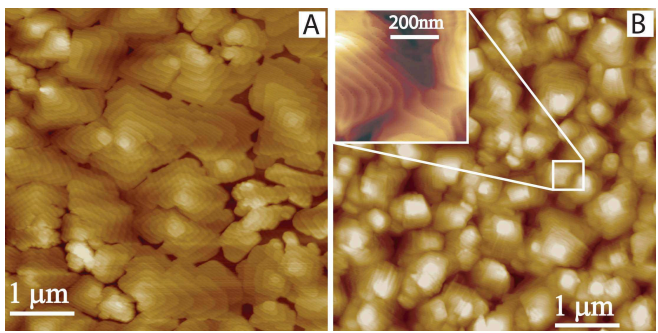}
\caption{AFM images of pentacene films grown on H-terminated Si
(A) and on glass (B). The inset of (B) shows the layered structure
of the grains for the films deposited on glass.}
  \label{fig:AFM}
\end{figure}

We attribute the difference in the electrical behavior observed
for films deposited on Si and glass substrates to the difference
in film morphology, which we have analyzed using an atomic force
microscope. Figure \ref{fig:AFM} shows AFM images of two pentacene
films of similar thickness deposited on the hydrogen terminated Si
(Fig.\ref{fig:AFM}A) and on the glass surface
(Fig.\ref{fig:AFM}B). It is apparent that very different
morphologies are observed for the two substrates. PEN films
deposited on Si surfaces exhibit large crystalline grains with a
common relative orientation and only relatively small fluctuations
in height. On glass, on the contrary, the grains are much smaller,
randomly oriented and they exhibit much larger height
fluctuations.\\

Our findings are consistent with the studies of Ruiz et al, which
had shown that the growth and morphology of pentacene films is
strongly influenced by the surface termination of the substrate
\cite{PENgrowth}. Specifically, for pentacene films grown on
SiO$_{2}$, a high density of nucleation centers was observed
leading to the growth of small islands and to a high concentration
of grain boundaries. On the hydrogen terminated silicon surface,
on the contrary, a much smaller density of nucleation centers was
observed resulting in significantly larger islands and in a
reduced density of grain boundaries. The critical influence of the
film morphology on the electrical characteristics of
electron-doped pentacene films is further supported by very recent
experiments \cite{Matsuo3} by Kanedo et al. in which metallic
conduction in rubidium-doped films deposited on glass was observed
after performing a high-temperature annealing on the doped films,
resulting in an improved morphological quality.\\

The most striking aspect of our observations, that had never been
previously reported, is the sharp suppression occurring as the
doping level is increased above 1 K/PEN. In this respect, PEN
behaves differently from C$_{60}$ \cite{Forro} and the metal
phthalocyanines \cite{MPcAdvMat,MPcJACS}, the only other molecular
compounds in which a metallic state has been induced by alkali
intercalation. For these molecular systems, a much larger number
of potassium atoms per molecule are needed to bring the film back
into an insulating state (six for C$_{60}$ and four for the metal
phthalocyanines). This observation is important, as it implies
that the conductivity of pentacene at high doping cannot be
described in terms of independent electrons filling the molecular
band originating from the LUMO orbital. In fact, for independent
electrons the conductivity at 1 K/PEN should be metallic -as
1K/PEN corresponds to a half filled band-and should remain
metallic up to 2 K/PEN, when all the spin-degenerate state in the
molecular band are completely filled. This is not what is
observed, as the films already enter an insulating state at
potassium concentration higher than 1.1 K/PEN.

Having excluded a picture founded on independent electrons, our
data point to an interpretation based on the concept of
Mott-Hubbard insulator \cite{MottHubard}. For most molecular
systems, the on-site Coulomb interaction U of two charge carriers
residing on the same molecule is much larger than the bandwidth W.
In such a regime, transport of electrons is blocked at half
filling because electron motion necessarily requires double
occupation of molecular sites that are not energetically
accessible. When this happens, the system is insulating due to
strong-correlations induced by electron-electron interaction. For
the oligoacenes (anthracene, tetracene, and pentacene) and the
olithophenes, recent \textit{ab-initio} calculations
\cite{electronic-corelations} show that U is always much larger
than W. In particular, for pentacene $W = 0.6 eV$ and $U = 1.4
eV$. This corresponds to a U/W ratio of 2.3, for which a
Mott-Hubbard insulating state is indeed expected and can therefore
explain the transition to an insulating state observed in the data
of Fig. \ref{fig:fig1}A.

This interpretation implies that the charge transfer from the
potassium atoms to the PEN molecules saturates at concentrations
higher than 1 K/PEN, since otherwise a metallic conductivity
should reappear when potassium concentration is increased well
beyond 1 K/PEN. Such a conclusion is consistent with existing
studies of charge transfer from alkali atoms to pentacene
\cite{spectroscopy1,spectroscopy2} in a solid state matrix, in
which only monoionic and no doubly charged states have been
observed. Thus at potassium concentrations higher than 1 K/PEN the
electron concentration in the films remains fixed at one electron
per molecule, corresponding to a half-filled band. In this
scenario, the width of the decay in conductivity observed for
$K/PEN >1$ (Fig.\ref{fig:fig1}A) is determined by the
inhomogeneity of the potassium concentration that, from our RBS
measurements (Fig.\ref{fig:fig1}C), is estimated to be at the 20
$\%$ level. These considerations make clear that the saturation of
charge transfer is essential for the observation of the
Mott-Hubbard state on the conductivity of non-stoichiometrically
doped films. In fact, as an insulating behavior due to
Mott-Hubbard physics is only visible in a narrow range of doping
close to one electron per molecule, the inhomogeneity in electron
concentration present in the absence of charge transfer saturation
would prevent its observation.

In conclusion, we have shown that in potassium-doped pentacene a
metallic state occurs in a broad range of concentrations up to
approximately 1 K/PEN. At this value, charge transfer from
potassium to PEN saturates, leaving the films at half filling.
This causes an insulating temperature dependence of the
conductivity that we attribute to a Mott-Hubbard insulating state
recently predicted from \textit{ab-initio} calculations
\cite{electronic-corelations}.

\end{document}